\documentstyle[epsf]{article}

\newcommand{\sss}{\scriptscriptstyle}

\begin{document}

\begin{titlepage}
\begin{flushright}MADPH-97-1030
\end{flushright} 
\begin{flushright}December, 1997
\end{flushright}
\vspace{2truecm}
\begin{center}
{\large\bf
Flavor Changing Neutral Currents \\ at $\mu^+\mu^-$ Colliders}\\
\vspace{.5in}
{\bf L.~Reina}\\
\vspace{.3in}
{\it  Physics Department, University of Wisconsin,\\
Madison, WI 53706, USA}
\vspace{1in}  
\end{center}
\begin{abstract} 
We illustrate the possibility of observing signals from Flavor
Changing Neutral Currents, originating from the scalar sector of a Two
Higgs Doublet Model. In particular, we focus on the tree level process
$\mu^+\mu^-\rightarrow \bar t c+\bar c t$, via scalar exchange in the
$s$-channel, as a distinctive process for $\mu^+\mu^-$ colliders.
\end{abstract}
\vspace{3truecm}
\begin{center}
To appear in the Proceedings of the \emph{Workshop on Physics at the
First Muon Collider and at the Front End of a Muon Collider}, Fermilab,
November 6-9, 1997.
\end{center}
\end{titlepage}
\clearpage 

\section*{Introduction}

The First Muon Collider (FMC) will reach a maximum center of mass
energy of 500 GeV, exploring all the interesting intermediate
regimes. In a second phase, the $\mu^+\mu^-$ collider should upgrade
its energy to up to 4 TeV and therefore become a very high energy
lepton collider, even compared to NLC.

A high energy lepton collider will be a very promising environment to
look for {\it new physics} beyond the Standard Model (SM), taking
advantage of the large enough energy which will become available at
low background rates. As has become common knowledge by now, the great
advantage of a $\mu^+\mu^-$ machine will be to allow the study of the
$s$-channel production of scalar \emph{Higgs like} particles, which
should occur at a much more conspicuous rate with respect to an
$e^+e^-$ collider ($m_\mu\!\simeq\! 200\,m_e$).

A very important and direct application of this property will be the
study of the Higgs sector of both the SM and its SuperSymmetric (SUSY)
extensions, looking for positive or negative evidence of the scalar
and pseudoscalar particles which are theoretically predicted in these
models. Even if evidence is found at a different machine (for instance
a hadron collider), a $\mu^+\mu^-$ machine will offer an optimal
energy resolution to sit at resonance and study the properties of
these particles, i.e. their masses and their couplings. This subject
has been thoroughly covered in some plenary talks
\cite{lreina:bargerfnal97,lreina:gunionfnal97} and dedicated parallel
sessions \cite{lreina:hanfnal97,lreina:carenafnal97} during this
workshop.

Along these lines, we want to discuss here the possibility of further
studying the properties of the scalar sector of the SM as well as of
many of its extensions, SUSY included, by looking for \emph{anomalous}
Flavor Changing Neutral Currents (FCNC) induced by scalar exchange.
It is well known that any extended scalar sector containing identical
replicas of the same representation of scalar fields (for instance: N
identical doublets or N identical pairs of doublets, etc.) can induce
FCNC at the tree level. This is due to the possibility of
diagonalizing the mass matrix of the fermion fields without
diagonalizing each single fermion-scalar coupling\footnote{We assume
Yukawa type couplings between fermions and scalars and fermion masses
generated through spontaneous symmetry breaking.}. The interest in
models containig many identical generations of scalars (sometimes even
in a one-to-one correspondence with the generations of fermions)
arises in many string-inspired Grand Unified Models (for instance
$E_6$) and is therefore of a more general interest.

In order to be more predictive and to limit the number of parameters
in our analysis, we focus on a minimal extension of the scalar sector
of the SM, a Two Higgs Doublet Model (2HDM). This can be used 
as a simple model, which we are able to work out to the very last
consequences in order to study the compatibility of its predictions
with the existing experiments.  On a more general ground, our analysis
should provide useful hints for a diversity of extensions of the SM,
which both theorist and experimentalist are encouraged to study.

Since FCNC are forbidden in the SM, their study can provide us with
unambiguous evidence of {\it new physics}. However, severe constraints
are imposed by the low energy physics of the $K$- and $B$-mesons, such
that FCNC have practically to be avoided in this sector of the
theory. This is naturally accomplished by the SM itself via the GIM
mechanism, and has to be imposed {\it ad hoc} in any 2HDM, by
introducing a discrete symmetry\cite{lreina:glash77}, which limits the
possible Yukawa couplings between fermions and scalars.

Apart from the experimental constraints coming from $K$- and
$B$-physics, there is no {\it a priori} theoretical reason not to have
FCNC. Therefore, the assumption of this {\it ad hoc} discrete symmetry
may be dropped in favor of a more natural one, which takes any Flavor
Changing (FC) coupling to a scalar field to be proportional to the
mass of the coupled fermions. The basic idea is that a natural
hierarcy is provided by the observed fermion masses and this may be
transfered to the couplings between fermions and scalar fields
\cite{lreina:sher87,lreina:sher91,lreina:hall92,lreina:hall93}. In
this way, FCNC are naturally suppressed in the light sector of the
theory, while dramatic effects may be seen in processes which involve
the heavy quark fields of the third generation, i.e. the top quark.

We illustrate these ideas at work in the following sections, first
presenting the model we refer to
\cite{lreina:savage93,lreina:atwoodetal97} and then focusing on some
FC signals, namely $(\bar t c+\bar c t)$-production which, if possible
at an $e^+e^-$-collider \cite{lreina:atwoodetal96}, is even more
enhanced and distinctive at a $\mu^+\mu^-$-collider
\cite{lreina:atwoodetal95}.

\section*{The Model}

We explicitly consider in this context only the quark fields, assuming
that the discussion of the quark and lepton sectors of the theory can
proceed independently. Then, let us consider the quark Yukawa
Lagrangian of a 2HDM, which we write as,

\begin{equation}
{\cal L}^{(III)}_{Y}= \eta^{U}_{ij} \bar Q_{i,L} \tilde\phi_1 U_{j,R} +
\eta^D_{ij} \bar Q_{i,L}\phi_1 D_{j,R} + 
\xi^{U}_{ij} \bar Q_{i,L}\tilde\phi_2 U_{j,R}
+\xi^D_{ij}\bar Q_{i,L} \phi_2 D_{j,R} \,+\, h.c. 
\label{E:lreina:1}
\end{equation}

\noindent where $\phi_i$, for $i=1,2$, are the two scalar doublets of
a 2HDM ($\tilde\phi_i\!=\!i\sigma^2\phi_i$), while $\eta^{U,D}_{ij}$
and $\xi_{ij}^{U,D}$ are the non diagonal matrices of the Yukawa
couplings.  

\noindent In order to prevent FCNC to arise at the tree level, the
scalar potential and Yukawa Lagrangian need to be constrained by an
{\it ad hoc\/} discrete symmetry \cite{lreina:glash77},

\begin{eqnarray}
\phi_1 \rightarrow -\phi_1 \,\,\,\,\,\,\,\,\,\mbox{and}&&
\,\,\,\,\,\phi_2\rightarrow\phi_2\\ \label{E:lreina:2}
D_i\rightarrow -D_i \,\,\,\,\,\,\,\,\,\mbox{and}&&\,\,\,\,\, 
U_i\rightarrow\mp U_i\,\,\,\,.\nonumber
\end{eqnarray}

\noindent Depending on whether the up-type and down-type quarks are
coupled to the same or to two different scalar doublets respectively,
one obtains the so called Model~I and Model~II 2HDM's.
\cite{lreina:hunter90}.

In contrast we want to consider the case in which no discrete symmetry
is imposed and both up-type and down-type quarks then have FC
couplings. For this type of 2HDM, that we call Model~III, the Yukawa
Lagrangian for the quark fields is as in Eq.~(\ref{E:lreina:1}) and no
term can be dropped {\it a priori}, see also
Refs.~\cite{lreina:savage93,lreina:atwoodetal96}. Since the two scalar
doublet are completely independent , by a suitable rotation of the
quark fields, we can chose the two scalar doublets in such a way that
only the $\eta_{ij}^{U,D}$ couplings generate the fermion masses,
i.e. such that

\begin{equation}
<\phi_1>\,=\left( 
\begin{array}{c}
0\\
{v/\sqrt{2}}
\end{array}
\right), \ \ \ \ 
<\phi_2>\,=0 \,\,\,\,.
\label{E:lreina:3}
\end{equation}

\noindent To the extent that the definition of the $\xi^{U,D}_{ij}$
couplings remains arbitrary, we will denote by $\xi^{U,D}_{ij}$ the
new rotated couplings, such that the charged couplings look like
$\xi^{U}\cdot V_{\sss{\rm CKM}}$ and $V_{\sss{\rm CKM}}\cdot\xi^{D}$.
This form of the charged couplings is indeed peculiar to Model~III
compared to Models~I and II and can have important phenomenological
repercussions \cite{lreina:atwoodetalrb96,lreina:atwoodetal97}.

\noindent The scalar physical mass spectrum consists of two charged
$\phi^{\pm}$ and three neutral spin 0 bosons, two scalars ($H^0,h^0$)
and a pseudoscalar ($A^0$),

\begin{eqnarray} 
H^0 & = & \sqrt{2}[(\mbox{Re}\,\phi^0_1-v) \cos \alpha+ \mbox{Re}\,
\phi^0_2 \sin \alpha ]
\nonumber \\ 
h^0 & = & \sqrt{2}[-(\mbox{Re}\,\phi^0_1-v) \sin\alpha +\mbox{Re}\,
\phi^0_2\cos\alpha ] \label{E:lreina:4} \\ 
A^0 & = & \sqrt{2} (-\mbox{Im}\,\phi^0_2)\,\,\,\,,
\nonumber 
\end{eqnarray}

\noindent where $\alpha$ is a mixing phase which also determines the
couplings between the neutral scalars and gauge bosons\footnote{We
remind that in a 2HDM the pseudoscalar field $A^0$ does not couple to
the gauge bosons.} ($W^\pm$, $Z^0$), i.e.

\begin{tabular}{p{6cm} p{5cm}}\\ \\ \\
\parbox[b]{6cm}{\epsffile{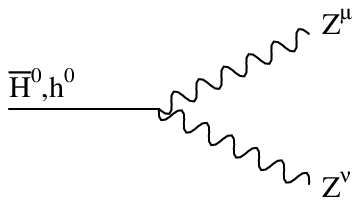}} & 
\raisebox{5.ex}{$\pm i\frac{g_{\sss W}}{c_{\sss W}}M_Z
(\cos\alpha,\,\sin\alpha)\, g^{\mu\nu}$}\\ \\ \\
\parbox[b]{6cm}{\epsffile{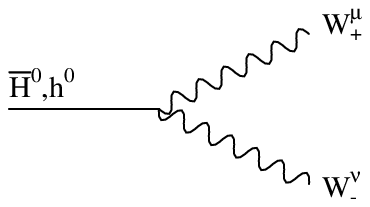}} & 
\raisebox{5.ex}{$\pm ig_{\sss W}M_W 
(\cos\alpha,\,\sin\alpha)\, g^{\mu\nu}\,\,\,\,.$}\\ \\ \\
\end{tabular}

\noindent It is interesting to notice that for $\alpha=0$: (\emph{1})
$H^0$ corresponds exactly to the SM Higgs field, and $\phi^{\pm}$,
$h^0$ and $A^0$ generate the new FC couplings; (\emph{2}) $h^0$ does
not couple to the gauge bosons, i.e. it behaves like the pseudoscalar
field $A^0$.

Finally, we want to introduce some definite ansatz that will guide us
in our phenomenological approach to the study of Model~III.  Because
the Yukawa Lagrangian directly breaks the flavor symmetry among
quarks, and this ultimately results into fermion mass generation, some
major proposals exist in the
literature\cite{lreina:sher87,lreina:sher91,lreina:hall92,lreina:hall93}
which suggest to take the new FC couplings to be proportional to the
mass of the quarks involved in the coupling, i.e.

\begin{equation} 
\xi_{ij} =
\lambda_{ij}\frac{\sqrt{m_im_j}}{v} \,\,\,\,,
\label{E:lreina:5}
\end{equation}

\noindent where for the sake of simplicity we take the $\lambda_{ij}$
to be real (for more details see
\cite{lreina:atwoodetal96,lreina:atwoodetal97}).
In this ansatz the residual degree of arbitrariness of the FC
couplings is expressed through the $\lambda_{ij}$ parameters, which
need to be constrained by the available phenomenology. In particular
we will see how $K^0\!-\!\bar K^0$ and $B^0\!-\!\bar B^0$ mixings (and
to a less extent $D^0\!-\!\bar D^0$ mixing) put severe constraints on
the FC couplings involving the first family of quarks. 

There is no doubt that the most interesting signals of these
non-standard couplings are to come from the physics of the top quark,
both production and decays. Therefore, we would like to single out the
right processes and the right environment in which we could already
have the possibility of testing the consequences of our assumptions.

\section*{Analysis of the constraints}

The existence of FC couplings is very much constrained by the
experimental results on $F^0\!-\!\bar F^0$ flavor mixings (for
$F\!=\!K,B$ and to a less extent $D$)

\begin{eqnarray}
\label{E:lreina:6}
\Delta M_K&\simeq& 3.51\cdot 10^{-15}\,\,\mbox{GeV}\nonumber\\
\Delta M_{B_d}&\simeq& 3.26\cdot 10^{-13}\,\,\mbox{GeV}\\ 
\Delta M_D&<& 1.32\cdot 10^{-13}\,\,\mbox{GeV}\,\,\,\,,\nonumber
\end{eqnarray}

\noindent due to the presence of new tree level contributions to each
of the previous mixings. We have analyzed the problem in detail in
Ref.~\cite{lreina:atwoodetal97}, taking into account both tree level
and loop contributions. Indeed the two classes of contributions can
affect different FC couplings, due to the peculiar structure of the
charged scalar couplings (see previous section).

We find that, unless for scalar masses in the multi-TeV range, the
tree level contributions need to be strongly suppressed, requiring
that the corresponding FC couplings are much than one. Enforcing the
ansatz made in Eq.~(\ref{E:lreina:5}), this amounts to demand that

\begin{equation}
\lambda^D_{ds}\ll 1\,\,\,\,,\,\,\,\,\lambda^D_{db}\ll 1\,\,\,\,
\mbox{and}\,\,\,\, \lambda^U_{ud}\ll 1\,\,.
\label{E:lreina:7}
\end{equation}

\noindent More generally, we can assume that the FC couplings
involving the first generation are negligible. Particular 2HDM's have
been proposed in the literature in which this pattern can be realized
\cite{lreina:kao96}. The remaining FC couplings, namely $\xi^U_{ct}$
and $\xi^D_{sb}$ are not so drastically affected by the $F^0\!-\!\bar
F^0$ mixing phenomenology. {}From the analysis of the loop
contributions to the $F^0\!-\!\bar F^0$ mixings (box and penguin
diagrams involving the new scalar fields), we verify that many regions
of the parameter space are compatible with the results in
Eq.~(\ref{E:lreina:6}) \cite{lreina:atwoodetal97}. Therefore we may
want to look at other constraints in order to single out the most
interesting scenarios.

Three are in particular the physical observables that impose strong
bounds on the masses and couplings of Model~III
\cite{lreina:atwoodetalrb96,lreina:atwoodetal97}
\begin{itemize}
\item the inclusive branching ratio for $B\rightarrow X_s\gamma$,
which is measured to be \cite{lreina:alam95}

\begin{equation}
BR(B\rightarrow X_s\gamma)=(2.32\pm 0.51\pm 0.29\pm 0.32)\times
10^{-4}\,\,\,\,,
\label{E:lreina:8}
\end{equation}

\item the ratio $R_b\!=\!\Gamma(Z\rightarrow b\bar
b)/\Gamma(Z\rightarrow \mbox{hadrons})$, whose present measurement
\cite{lreina:lepwg97} is such that $R_b^{\rm expt}>R_b^{\rm SM}$
($\sim 1.4\sigma$),

\begin{equation}
R_b^{\rm expt}= 0.2170\pm 0.0009 \label{E:lreina:10} \,\,\,\,\,
\mbox{while} \,\,\,\,\,
R_b^{\rm SM}=0.2158\,\,\,\,, 
\end{equation}

\item the corrections to the $\rho$ parameter, which has become
conventional to describe in terms of

\begin{equation}
\rho_0=\frac{M_W^2}{\rho M_Z^2\cos^2\theta_W} =
1+\Delta\rho_0^{\sss{\rm NEW}} \,\,\,\,
\label{E:lreina:11}
\end{equation}

\noindent where $\rho$ absorbs all the SM corrections to the gauge
boson self energies and, in the presence of new physics,
$\Delta\rho_0^{\sss{\rm NEW}}$ summarizes the deviation from the SM
prediction (i.e. $\rho_0\!=\!1$). {}From the recent global fits of the
electroweak data, which include the input for $m_t$ from
Ref.~\cite{lreina:giromini97} and the new experimental results on
$R_b$, $\rho_0$ turns out to be very close to unity. This imposes
severe constraints on many extension of the SM, especially on the mass
range of the new particles.
\end{itemize}

Since the experimental determination of $R_b$ is not completely
definite, we demand less strict agreement between $R_b^{\rm exp}$ and
$R_b^{\rm SM}$. In this case we find compatibility with the present
experiments for

\begin{equation}
\lambda_{ct}\simeq O(1)\,\,\,\,\mbox{and}\,\,\,\,
\lambda_{sb}\simeq O(1)\,\,. 
\label{E:lreina:12}
\end{equation}

\noindent The value of the mixing angle $\alpha$ is not determinant,
while the masses are mainly dictated by the fit to $Br(B\rightarrow
X_s\gamma)$ and $\Delta\rho_0$ \cite{lreina:atwoodetalrb96}. We are
left with two possible scenarios,

\begin{equation}
M_H,M_h\le M_c\le M_A\,\,\,\,\,\mbox{or}\,\,\,\,
M_A\le M_c\le M_H,M_h \,\,\,.
\label{E:lreina:13}
\end{equation}

We conclude that, given the existing constraints, Model~III has the
very interesting characteristic of providing sizable FC couplings for
the top quark, in a way that will certainly be testable at the next
generation of lepton and hadron colliders. We will discuss some of
these phenomenological issues in the next section.

\section*{Top-charm production: the case of a muon collider}

If we assume Eq.~(\ref{E:lreina:12}), $\xi^U_{ct}$ becomes the most
relevant FC coupling. The presence of a $\xi^U_{ct}$ flavor changing
coupling can be tested by looking at both top decays and top
production (see Ref.~\cite{lreina:atwoodetal97} and references
therein).

It is interesting that the first upper bounds on $t\rightarrow c V$
($V\!=\!\gamma,Z^0$) are now coming from recent experimental analysis
\cite{lreina:tcVexp97} (see Ref.~\cite{lreina:atwoodetal97} for the
corresponding theoretical prediction in Model~III). Encouraged by this
progress, we want to concentrate here on top-charm production.  The SM
prediction for top-charm production is extremely suppressed and any
signal would be a clear evidence of new physics with large FC
couplings in the third family.  The final state for this process has a
unique kinematics, with a very massive jet recoiling against an almost
massless one (very different from a $bs$ production signal, for
instance). This quite distinctive signature may allow to work even
with relatively low statistics, as can be the case for a lepton
collider. The much better statistics one could get at an hadron
collider, would come at a cost of a much higher background (mostly,
tree level SM background for a one-loop process). A nice analysis of
the hadron collider case is presented in Ref.~\cite{lreina:peccei95}.

Here we want to focus on lepton colliders and in particular on the
case of the FMC.  In principle, the production of top-charm pairs
arises both at the tree level, via the $s$-channel exchange of a
scalar field with FC couplings, and at the one loop level, via
corrections to the $Ztc$ and $\gamma tc$ vertices.  

The $s$-channel production is not relevant for an $e^+e^-$ collider,
because the coupling of the scalar fields to the electron is very
suppressed ($m_\mu\simeq 200 \,m_e$).  An interesting proposal for a
$t$-channel production via $W^+W^-$-fusion at a $\sqrt{s}\!=\!1$ TeV
$e^+e^-$ collider has been pointed out in Ref.~\cite{lreina:soni97}.
However, top-charm production at an $e^+e^-$ collider remains mainly a
loop effect and therefore (even at the energies of NLC) it is
suppressed with respect to the corresponding production cross section
at the FMC (see Ref.~\cite{lreina:atwoodetal96}) and in particular at
the very high energy muon collider.

In fact, for a $\mu^+\mu^-$ collider, the $s$-channel top-charm
production via a neutral scalar/pseudoscalar is a nice example of the
kind of resonance production we briefly discussed in the Introduction.
At resonance, the new scalars (let us denote them generically by $\cal
H$) may be produced at an appreciable rate, and the effective cross
section for any final state very much depends on the relation between
the beam energy resolution and the width of the scalar particle
($\Gamma_{\cal H}$) produced in the $s$-channel
\cite{lreina:barger95}.

\noindent Following \cite{lreina:barger95}, we define the effective
cross section as the convolution of the Breit-Wigner
$\sigma^{BW}_{tc}$ cross section with a gaussian beam energy spread,

\begin{equation}
\sigma^{eff}_{tc}=\int d\sqrt{s^\prime}
\frac{\mbox{exp}[-(\sqrt{s^\prime}-\sqrt{s})^2/2\sigma^2]}
{\sqrt{2\pi}\sigma}\,\sigma^{BW}_{tc}(s^\prime)\,\,\,\,,
\label{E:lreina:14}
\end{equation}

\noindent where the rsm of the gaussian distribution is defined in
terms of the \emph{resolution} parameter $R$ as

\begin{equation}
\sigma = 7\,\mbox{MeV} 
\left(\frac{R}{0.01}\right) 
\left(\frac{\sqrt{s}}{100\,\mbox{GeV}}\right)\,\,\,\,.
\label{E:lreina:15}
\end{equation}

\noindent If $\sigma\gg\Gamma_H$ the effective cross section is
suppressed as $\Gamma_{\cal H}/s$, 
\begin{equation}
\sigma^{eff}_{tc}=\frac{\pi\Gamma_H}{2\sqrt{2\pi}\sigma}
\sigma_{BW}(s=M_{\cal H})\,\,\,\,,
\label{E:lreina:16}
\end{equation}
\noindent while the optimal case is reached if $\sigma\ll\Gamma_H$,
when
\begin{equation}
\sigma^{eff}_{tc}=\sigma_{BW}(s=M_{\cal H})\,\,\,\,,
\label{E:lreina:17}
\end{equation}
\noindent with a whole spectrum of possible intermediate cases. In our
analysis we study
\begin{equation}
R_{tc}=\frac{\sigma^{eff}_{tc}}{\sigma_0}=
R({\cal H})\left(B({\cal H}\rightarrow \bar t c) + 
B({\cal H}\rightarrow\bar c t)\right)\,\,\,\,,
\label{E:lreina:18}
\end{equation}
\noindent where
$\sigma_0\!=\!\sigma(\mu^+\mu^-\rightarrow\gamma\rightarrow e^+e^-)$
and $R({\cal H})\!=\!\sigma_{\cal H}/\sigma_0$ for $\sigma_{\cal H}$
the total cross section for producing ${\cal H}$. 

To be more explicit, let us consider the case of a scalar
field ${\cal H}=h^0$. Then, according to what we discussed in a
previous section,

\begin{equation}
C_{h^0tc}= \frac{1}{\sqrt{2}}\left[ \xi_{tc}P_R+\xi_{ct}^\dagger P_L 
\right]
\cos\alpha \equiv \frac{g\sqrt{m_t m_c}}{2 M_W}(\chi_R P_R+\chi_L P_L)
\,\,\,\,.
\label{E:lreina:19}
\end{equation}

The total width $\Gamma_{h^0}$ can be obtained from the literature
(see for instance \cite{lreina:hunter90}), and varies with $M_{h^0}$
because of the different decay channels that open up at higher
$M_{h^0}$. On the other hand, the rate for top-charm production in
Model~III is explicitly given by,
\begin{equation}
\Gamma(h^0\rightarrow t\bar c)=
\frac{3 g^2 m_t m_c M_{h^0}}{32\pi M_W^2} 
\left( \frac{(M_{h^0}^2-m_t^2)^2}{M_{h^0}^4} \right) 
\left({|\chi_R|^2+|\chi_L|^2\over 2} \right)\,\,\,\,.
\label{E:lreina:20}
\end{equation}

\noindent Therefore, our results depend on both $M_{h^0}$ and
$\Gamma_{h^0}$, and on how $\Gamma_{h^0}$ compares to the resolution
parameter $R$. We use the set of parameters posted for this
workshop. In particular we consider the possibility to reach
resolutions as accurate as $R\!\simeq\!1\%$ and the availability of a
average luminosity of ${\cal L}_{\mbox{av}}\!=\!10^{32}$ for
$\sqrt{s}\!=\!  200$ GeV up to ${\cal L}_{\mbox{av}}\!=\!7\times
10^{32}$ for $\sqrt{s}\!=\!  500$ GeV.

\begin{figure}[ht]
\centering
\epsfxsize=5.in
\leavevmode\epsffile{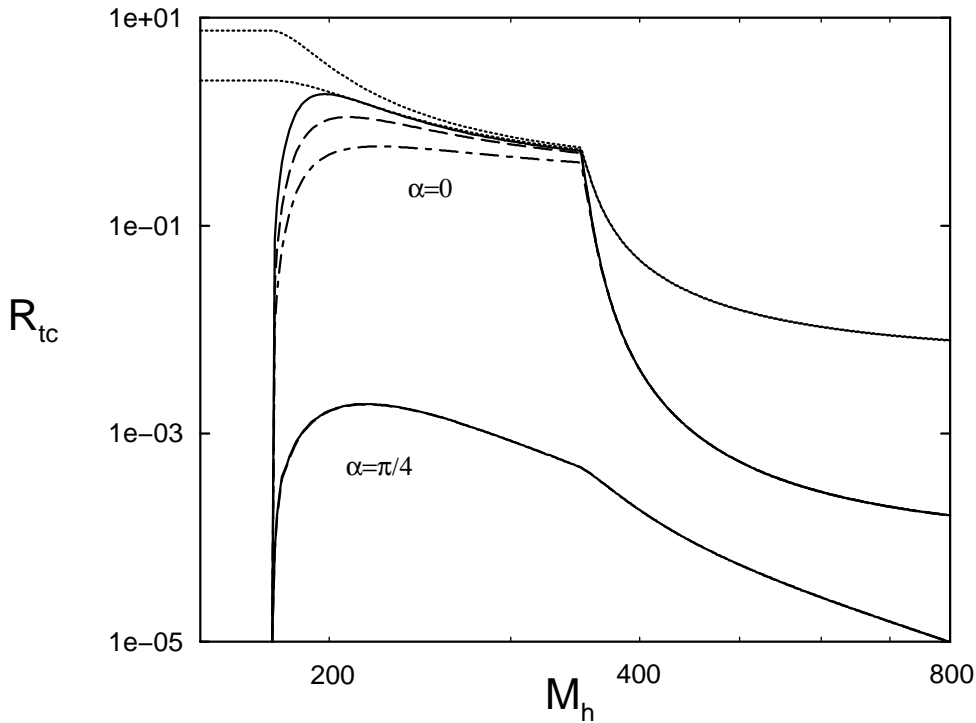}
\caption[]{The value of $R(h^0)$ is shown as a function of $M_{h^0}$
in a pure Breit-Wigner case (upper dotted line) and when the gaussian
width distribution of the beam energy (for $R\!=\!0.01$) is taken into
account (lower dotted line).  The ratio $R_{tc}$ is also shown for
different values of the resolution parameter $R\!=\!0.001$ (solid),
0.01 (dashed) and 0.03 (dot-dashed), when $\alpha\!=\!0$ (upper group
of curves) and when $\alpha\!=\!\pi/4$ (lower group of curves).}
\label{F:lreina:1}
\end{figure}

\noindent For a given $R$, we vary $M_{h^0}$ in the range,

\begin{equation}
100\,\mbox{GeV}\le M_{h^0}\le 800\,\mbox{GeV}\,\,\,\,,
\label{E:reina:21}
\end{equation}

\noindent and for given $M_{h^0}$ and $R$, we consider two different
cases: $\alpha\!=\!0$ and $\alpha\!=\!\pi/4$.  In the first case
$\Gamma_{h^0}$ is smaller than in the second case, because the $h^0$
field does not couple to $W^+W^-$ and $Z^0Z^0$. In both cases we
assume all the $\chi_i$ couplings to be real and of $O(1)$.  Our
results are summarized in Fig.~\ref{F:lreina:1}.

\noindent Note that for $\alpha\!=\!0$, if $M_{h^0}$ is below the
$t\bar t$ threshold $R_{tc}$ is about $.2-3$ and in fact $tc$ makes up
a large branching ratio.  Above the $t\bar t$ threshold $R_{tc}$
drops. In this \emph{narrow width} case we clearly see how our result
depends on $R$ and how for smaller values of $R$ the predictions get
closer and closer to the pure Breit-Wigner case. For
$\alpha\!=\!\pi/4$ the branching ratio is smaller due to the $W^+W^-$
and $Z^0Z^0$ threshold at about the same mass as the $tc$ threshold
and so $R_{tc}$ is around $10^{-3}$. In this case the width is
\emph{broader} and there is almost no dependence on $R$. To be more
specific, let us assume that $M_{h^0}=300$~GeV, then $\sigma_0\approx
1$~pb.  For ${\cal L}_{\mbox{av}}\!=\!10^{32}$ cm$^{-2}$ s$^{-1}$ and
$R\!=\!0.01$, $\alpha\!=\!0$ will produce about $10^2\, (t\bar c+\bar
t c)$ events and $\alpha\!=\!\pi/4$ will produce only a few
events. Much higher statistics can be obtained improving on the
average luminosity available, which we hope will remain one of the
priorities in the study of the FMC. Given the distinctive nature of
the final state and the lack of a Standard Model background,
sufficient luminosity should allow the observation of such events.

If such events are observed, we will also have the possibility to 
extract the values of $\chi_L$ and $\chi_R$, provided we determine 
the helicity of the produced top quark, expressed in termes of
$\chi_L$ and $\chi_R$ as

\begin{equation}
{\bf H}_t=-{\bf H}_{\bar t}= \frac{|\chi_R|^2-|\chi_L|^2} 
{|\chi_R|^2+|\chi_L|^2}\,\,\,\,.
\label{E:lreina:22}
\end{equation}

\noindent The helicity of the $t$ quark cannot be determined directly,
but has to be obtained from the decay distributions of the top
\cite{lreina:soni95}, the number of events required to observe it with
a significance of $3\,\sigma$ being

\begin{equation}
N_{3\sigma}=\frac{36}{{\cal E}_t^2{\bf H}_t^2} \approx 
\frac{107} {{\bf H}_t^2}\,\,\,\,.
\label{E:lreina:23}
\end{equation}

\noindent Thus at least $10^2$ events are required to begin to measure
the helicity of the top and hence the relative strengths of $\chi_L$
and $\chi_R$.  In the above numerical examples it is clear that for
some combinations of parameters, particularly if the luminosity is
higher than $10^{32}$ cm$^{-2}$ s$^{-1}$, sufficient events to measure
the helicity may be present.

\section*{Acknowlegement}
This research was supported by the U.S. Department of Energy under
Contract DE-FG02-95ER40896.

\end{document}